\documentclass[amsmath,amssymb,amsfonts,twocolumn]{revtex4}
\usepackage{graphicx}
\usepackage{bbm}
\usepackage{braket,bigints}
\usepackage{amsmath,stmaryrd,mathtools}
\def \beq {\begin{equation}}
\def \eeq {\end{equation}}
\def \tr {\rm Tr}
\newcommand {\qs} {{\rm Q}_{\rm S}}
\newcommand {\qt} {{\rm Q}_{\rm T}}
\newcommand {\ks} {k_{\rm S}}
\newcommand {\kt} {k_{\rm T}}

\begin{document}
\title{Quantum information processing in the radical-pair mechanism: Haberkorn's theory violates the Ozawa entropy bound} 
\author{K. Mouloudakis$^{1}$}
\author{I. K. Kominis$^{1,2}$}
\email{ikominis@physics.uoc.gr}
\affiliation{$^1$Department of Physics, University of Crete, 70013 Heraklion, Greece\\
$^2$Institute for Theoretical and Computational Physics, University of Crete, 70013 Heraklion, Greece}

\begin{abstract}
Radical-ion-pair reactions, central for understanding the avian magnetic compass and spin transport in photosynthetic reaction centers, were recently shown to be a fruitful paradigm of the new synthesis of quantum information science with biological processes. We here show that the master equation so far constituting the theoretical foundation of spin chemistry violates fundamental bounds for the entropy of quantum systems, in particular the Ozawa bound. In contrast, a recently developed theory based on quantum measurements, quantum coherence measures and quantum retrodiction, thus exemplifying the paradigm of quantum biology, satisfies the Ozawa bound as well as the Lanford-Robinson bound on information extraction. By considering Groenewold's information, the quantum information extracted during the reaction, we reproduce the known and unravel new magnetic-field effects not conveyed by reaction yields.
\end{abstract}
\maketitle
\section{Introduction}
Quantum biology \cite{plenio_review,qubio_book,kominis_review} has been recently emerging as an interdisciplinary field pointing to certain biological processes which, counterintuitively, exhibit quantum coherent dynamics, and accordingly require for their understanding physical concepts developed in quantum information science. This is surprising since decoherence is ordinarily expected to be prevalent in complex biological matter. Yet, there appear to be several cases where decoherence is not as detrimental, and moreover, where quantum coherent dynamics seem to have an operational significance.

Prominent among such examples have been the excitation energy transport in photosynthetic light harvesting \cite{Van,Engel,Panitchayangkoon,Collini,Mosheni,Caruso} and the radical-pair mechanism \cite{kominis_review,komPRE2009,komPRE2011,komPRE2014,lamb,cpl2015,sun,plenioPRL,shao,kais}, which was introduced in the late 1960's \cite{Matysik_review} to explain unexpectedly large signals in NMR measurements of organic radicals.
The mechanism is the cornerstone of spin chemistry, a field of physical chemistry and photochemistry studying the effects of electron and nuclear spins in chemical reactions \cite{steiner}. Radical-pair (RP) reactions have been studied extensively because, besides their potential role in avian magnetoreception \cite{schulten,ritz,ww}, they regulate spin transport in photosynthetic reaction centers \cite{Matysik_review}.  

The radical-pair mechanism has recently attracted renewed attention when it was suggested \cite{komPRE2009} that RP reactions involve quantum measurement dynamics 
and require for their understanding concepts like quantum coherence measures and the quantum communications concept of quantum retrodiction \cite{komPRE2011,komPRE2014}, rendering the mechanism a vivid paradigm for quantum biology on the qualitative level. On the quantitative level, we have developed a new master equation describing the fundamental quantum dynamics of RP reactions \cite{kominis_review}, which departs from the traditional theory, attributed to Haberkorn \cite{Haberkorn}. 
Abstract or not \cite{Hore_Review}, the master equation describing the time evolution of the radical-pair spin density matrix is the starting point for virtually all theoretical predictions relevant to the radical-pair mechanism.

Yet, there currently is no consensus on which is the fundamental master equation describing radical-pair reactions. Moreover, it is tactically close to impossible to rule out one of the two contending theories, Haberkorn's and ours, based either on how sound their derivation is perceived to be, or by comparing absolute predictions for various observables. Therefore, we have recently resorted to testing the consistency of each theory on its own. Whereas consistency does not prove adequacy, the opposite is true. To this end, we have shown that when casting the theories into the perspective of single-RP quantum trajectories, Haberkorn's theory produces severe inconsistencies when comparing quantum-trajectory averages with the master equation. 

We here utilize formal inequalities pertaining to the entropy of quantum systems to conclusively demonstrate that Haberkorn's theory cannot stand as a fundamental master equation accounting for RP quantum dynamics. In particular, the Ozawa bound \cite{Ozawa} states that the average entropy of the outcomes of a quantum measurement can be at most equal to the entropy of the pre-measurement quantum state. This is why measurements convey information. While Haberkorn's theory is shown to violate this inequality, our newly developed theory satisfies it, along with yet another bound, the Lanford-Robinson \cite{Lanford} inequality limiting the maximum possible information extraction. We thus put recent discussions of this particular front of quantum biology on a firmer ground, enabling the full exploration of the relevant biological phenomena allowed by the underlying quantum dynamics.

In Sec. II we briefly describe the radical-pair mechanism and its biological significance, and reiterate the two contending theories, Haberkorn's and ours, attempting to account for the fundamental quantum dynamics of RP reactions. In Sec. III we present the test of the Ozawa bound, demonstrating its violation by Haberkorn's theory and its satisfaction by our theory, which also satisfies the Lanford-Robinson bound. The latter directly leads to the concept of extracted information, called Groenewold information. In Sec. IV we show that this abstract quantity can predict magnetic-field effects. Additionally to the known ones, we show that Groenewold information can predict a new magnetic-field effect not conveyed by reaction yields.
\section{Quantum dynamics of radical-pair reactions}
Radical-ion pairs are biomolecular ions (each carrying an unpaired electron $\bullet$) created by an electron transfer from a photo-excited donor-acceptor molecular dyad: ${\rm DA}\xrightarrow{h\nu} {\rm D^{*}A}\rightarrow {\rm D}^{\bullet +}{\rm A}^{\bullet -}$. The magnetic nuclei of D and A couple to the respective unpaired electron via hyperfine interactions, leading to singlet-triplet (S-T) mixing, i.e. a coherent oscillation of the spin state of the electrons and concomitantly the nuclear spins, denoted by $^{\rm S}{\rm D}^{\bullet +}{\rm A}^{\bullet -}\leftrightharpoons~^{\rm T}{\rm D}^{\bullet +}{\rm A}^{\bullet -}$. The reverse charge transfer, called charge recombination, terminates the reaction and spin-selectively leads to the formation of either singlet or triplet neutral reaction products. The theoretical description of RP reactions is accounted for by the density matrix $\rho$ describing the spin state of the molecule's two electrons and any number of present magnetic nuclei \cite{basis}. 

The time evolution of $\rho$ has traditionally been described \cite{Haberkorn} by Haberkorn's master equation (HME), ${{d\rho}/ {dt}}=-i[{\cal H},\rho]-\ks\big(\qs\rho+\rho\qs\big)/2-\kt\big(\qt\rho+\rho\qt\big)/2\label{Hab}$.
The first term is the ordinary unitary evolution driven by the intramolecule magnetic interactions contained in the Hamiltonian ${\cal H}$ (Zeeman, hyperfine etc). As singlet and triplet states are not eigenstates of ${\cal H}$, this term generates S-T coherence. The spin-dependent population loss of RPs is described by the other two terms, called reaction terms, involving two operators and two rates. The orthogonal projectors $\qs$ and $\qt$ project the RP spin state onto the electron singlet and triplet subspace, respectively. It is $\qs\qt=\qt\qs=0$ and $\qs+\qt=\mathbbmtt{1}$, where $\mathbbmtt{1}$ is the unit operator. The rates $\ks$ and $\kt$ are the singlet and triplet recombination rates, at which the RP population decays in a spin-selective way. 

The master equation developed by us \cite{komPRE2009,komPRE2011,komPRE2014} reads
\begin{align}
{{d\rho}\over {dt}}=&-i[{\cal H},\rho]\label{t1}\\  
&-{{k_{\rm S}+k_{\rm T}}\over 2}\big(\rho {\rm {\rm Q_S}}+{\rm {\rm Q_S}}\rho-2{\rm {\rm Q_S}}\rho {\rm {\rm Q_S}}\big)\label{t2}\\
&-(1-p_{\rm coh})\big(k_{\rm S}{\rm {\rm Q_S}}\rho {\rm {\rm Q_S}}+k_{\rm T} {\rm Q_T} \rho {\rm Q_T}\big)\label{t3}\\
&-p_{\rm coh}{{dr_{\rm S}+dr_{\rm T}}\over {dt}}{1\over {\tr\{\rho\}}}
\Big(\qs\rho\qs+\qt\rho\qt+\nonumber\\&\hspace{1in}{1\over p_{\rm coh}}\qs\rho\qt+{1\over p_{\rm coh}}\qt\rho\qs\Big).\label{t4}
\end{align}
S-T coherence is generated by ${\cal H}$, dissipated by the Lindblad term \eqref{t2}, which we formally derived in \cite{komPRE2009,lamb}, and quantified \cite{komPRE2014,plenio_coherence} by $p_{\rm coh}=p_{\rm coh}\llbracket\rho\rrbracket$, which is a map of the density matrix onto the real interval [0,1]. At the single-RP level, the term \eqref{t2} is translated into unobserved randomly occurring projections of the RP state $\rho$ to either the singlet RP state $\qs\rho\qs/{\rm Tr}\{\rho\qs\}$ with probability $dp_{\rm S}=(\ks+\kt)dt\tr\{\rho\qs\}/2$, or to the triplet RP state $\qt\rho\qt/{\rm Tr}\{\rho\qt\}$ with probability $dp_{\rm T}=(\ks+\kt)dt\tr\{\rho\qt\}/2$. The terms \eqref{t3} and \eqref{t4} are the reaction terms, reducing the RP population, given by $\tr\{\rho\}$, in a spin-selective way, and derived in \cite{komPRE2014} using the theory of quantum retrodiction. For both theories, the fraction of the RP population recombining into singlet and triplet neutral products within the interval $dt$ is $dr_{\rm S}=\ks dt\tr\{\rho\qs\}$ and $dr_{\rm T}=\kt dt\tr\{\rho\qt\}$, respectively. 

HME follows from our theory by {\it forcing} $p_{\rm coh}$ to be zero at all times, i.e. HME is a limiting case of our theory valid in the regime of strong spin relaxation, where S-T coherence decays in a time scale much faster than $1/(\ks+\kt)$. To avoid a misunderstanding, we note that $p_{\rm coh}$ {\it is not} a free parameter of our theory. It is a well-defined function of the density matrix $\rho$, which in turn is governed by the master equation \eqref{t1}-\eqref{t4} (see \cite{kominis_review} for the definition of $p_{\rm coh}$). Hence forcing the value $p_{\rm coh}=0$ at all times is, according to our theory, an unphysical situation, since the master equation \eqref{t1}-\eqref{t4} in general predicts a non-zero value of $p_{\rm coh}$ along the reaction. On the other hand, if there is some external relaxation mechanism damping S-T coherence at a rate higher than the intrinsic decay of $p_{\rm coh}$ due to term \eqref{t2}, then $p_{\rm coh}$ will rapidly (compared to the reaction time) approach zero, and both our theory and HME will predict qualitatively similar dynamics. But then again, S-T coherence will have truly dissipated. We will revisit this point in the following.

The difference between the two contending theories can be best elucidated by considering single-RP quantum trajectories. We recently undertook such an analysis \cite{cpl2015}. If one assumes, as has been the intuitive understanding in spin chemistry, that RPs only undergo unitary evolution until they recombine, then HME qualitatively and quantitatively disagrees with the quantum-trajectory average. However, one could decide to question the results of this quantum trajectory analysis, as was done in \cite{Comment_Jeschke}, albeit unsuccessfully \cite{Reply_Kominis}. The following provides an independent demonstration of HME's failure based on the entropy of quantum systems.
\subsection{The biological significance of understanding the quantum dynamics of the radical-pair mechanism}
For completeness, we briefly elaborate on the potential usefulness of the following discussion for the biological context of the subject matter.  The radical-pair master equation is central in the theoretical understanding of (i) the avian magnetic compass, and (ii) the spin transport effects in photosynthetic reaction centers. It is clear that having a fundamentally sound theory, or at the least, having a set of tools that test the physical acceptability of any theory purporting to be fundamental, is critical for making progress in these two fields. In particular, regarding (i) there have been several studies \cite{Vedral,Briegel,Chia} of the effect of decoherence on the avian compass. The results of those studies are qualitatively and quantitatively dependent on our understanding of the fundamental quantum dynamics of the radical-pair mechanism. So far these studies have been based on HME. Regarding (ii), there is a large amount of experimental data on CIDNP (chemically induced dynamic nuclear polarization) concerning non-equilibrium nuclear spin polarization produced by the radical-pair dynamics in photosynthetic reaction centers. Understanding these data, which e.g. are used to extract molecular structure information for the reaction centers, requires the understanding of the fundamental quantum dynamics of radical-pairs. For example, there are cases \cite{kominis_cidnp} where HME predicts a zero CIDNP signal, whereas our theory predicts a non-zero signal. As another example \cite{cpl2015}, at high magnetic fields HME predicts a CIDNP enhancement different by 15\% from our theory's prediction, and the difference could become significantly larger for different Hamiltonians. When precision becomes an issue, understanding the underlying theory of the radical-pair mechanism should be of importance. 
\section{Testing the Ozawa and Lanford-Robinson entropy bounds}
The Ozawa bound is comprehensively introduced in \cite{Jacobs}. Consider a system described by the density matrix $\rho$. If an efficient quantum measurement is performed, with the possible post-measurement states $\tilde{\rho}_n$ occurring with probabilities $p_n$, then
\beq
\sum_{n}p_n{\cal S}\llbracket\tilde{\rho}_n\rrbracket\leq {\cal S}\llbracket\rho\rrbracket\label{ozawa}
\eeq
where ${\cal S}\llbracket r\rrbracket=-{\rm Tr}\{r{\rm Log}(r)\}$ is the von-Neumann entropy of the density matrix $r$. The interpretation is that our ignorance after the measurement averaged over the possible measurement results, $\sum_{n}p_nS(\tilde{\rho}_n)$, should be smaller than our initial ignorance about the system, $S(\rho)$, if we are to extract information about the system. 
Before proceeding to test whether the two contending master equations satisfy the Ozawa bound, we need to lay out some prerequisites.
\subsection{Equal versus unequal recombination rates}
The case of RPs with equal recombination rates is particularly innocuous. Setting $\ks=\kt=k$ in HME, it is trivial to formally solve HME, the solution is $\rho=e^{-kt}R$, where $dR/dt=-i[{\cal H},R]$. Thus $R$ is a trace-preserving density matrix evolving unitarily. That is, the density matrix $R$ undergoes a physically acceptable evolution, while the RP density matrix $\rho$ describes the exact same physical state as $R$ but having an exponentially decaying population. Similarly, our master equation leads again to $\rho=e^{-kt}R$, but now $R$ satisfies the Lindblad equation $dR/dt=-i[{\cal H},R]-k(\qs R+R\qs-2\qs R\qs)$, which again is a physically acceptable law \cite{Breuer} of evolving a density matrix. Hence, in the special case of equal recombination rates, $\ks=\kt$, both theories produce physically acceptable evolution laws, so we do not expect any of them to violate any entropy bounds, and indeed, they do not. 
The problems with HME arise as soon as $\ks\neq\kt$. The parameter regime of unequal recombination rates is physically very interesting as it appears in several realistic cases, like photosynthetic reaction centers \cite{Matysik_review} or avian compass models \cite{Hore_Nature}. In the following we will focus on the case $\ks\neq\kt$, demonstrating that in this regime HME leads to a physically unacceptable behavior.
\subsection{State purity}
We will illustrate the violation of the entropy bounds by HME using two exemplary cases. Altough just one particular failure of the theory is enough to rule it out, we use two examples of such failure because the first necessitates an additional term in the master equation that could be perceived as weakening our argument. {\bf (B1)} In particular, if we use as initial state a pure singlet state, it is seen that HME keeps the evolved state pure at all times. This can be easily proved by showing that all time-derivatives of $\pi(t)=\tr\{\rho^2\}/\tr\{\rho\}^2$ are zero when evaluated at $t=0$. This feature of HME is problematic in its own right \cite{kominis_review}, and translates into zero pre- and post-measurement entropy. In contrast, our theory evolves an initially pure state into a mixture due to the dephasing term \eqref{t2}. To be able to compare the two theories, we add {\it to both theories} a small spin-randomizing (and trace-preserving) term of the form $-\gamma(\rho-{{\tr\{\rho\}}\over {\tr\{\mathbbmtt{1}\}}}\mathbbmtt{1})$, where the rate $\gamma$ is orders of magnitude smaller than all other pertinent rates. This term is a {\it physically acceptable} form of spin relaxation pushing $\rho$ towards a fully mixed density matrix, and both theories should produce physically meaningful results even with this relaxation term included. We will show that this is not the case for HME. Additionally, this relaxation term will be useful in demonstrating when HME is an adequate approximation to RP quantum dynamics. {\bf (B2)} As a second example, and in order to avoid the introduction of the spin-randomizing term and make our argument more robust, we start with the initial state being a mixed triplet state of the form $\qt/{\rm Tr}\{\qt\}$. Although unusual chemically, it has been considered \cite{Hore_PRL} in the context of entanglement in RP reactions. Physically the triplet initial state is as acceptable as the singlet. Again, the same violation of the entropy bounds results from HME. Before demonstrating the results of the test, we lay out the calculation of the pre- and post-measurement entropies.
\subsection{Pre-measurement and post-measurement entropies}
The idea behind the Ozawa bound test is the following. At at any given time, the radical-pair density matrix is "produced" by the master equation under consideration. Furthermore, at any given time, the number of singlet and triplet neutral products of the reaction, given by $dr_{\rm S}$ and $dr_{\rm T}$, respectively (defined in Sec. II), depends on the radical-pair density matrix at that time. This number constitutes a measurement probability, since the radical-pairs being in some state described by $\rho$ end up in the singlet or triplet projections $\rho_{\rm S}=\qs\rho\qs/{\rm Tr}\{\rho\qs\}$ and $\rho_{\rm T}=\qt\rho\qt/{\rm Tr}\{\rho\qt\}$, respectively, which describe the spin state of the neutral products. We can thus define the pre-measurement entropy, which is the entropy of the radical-pair state $\rho$, and the post-measurement entropy consisting of the entopies of the neutral product states. If the master equation producing the state $\rho$ is "unphysical", we expect these entropies will not be properly related, e.g. as dictated by the Ozawa bound. 
\begin{figure*}[t!]
\begin{center}
\includegraphics[width=13.5 cm]{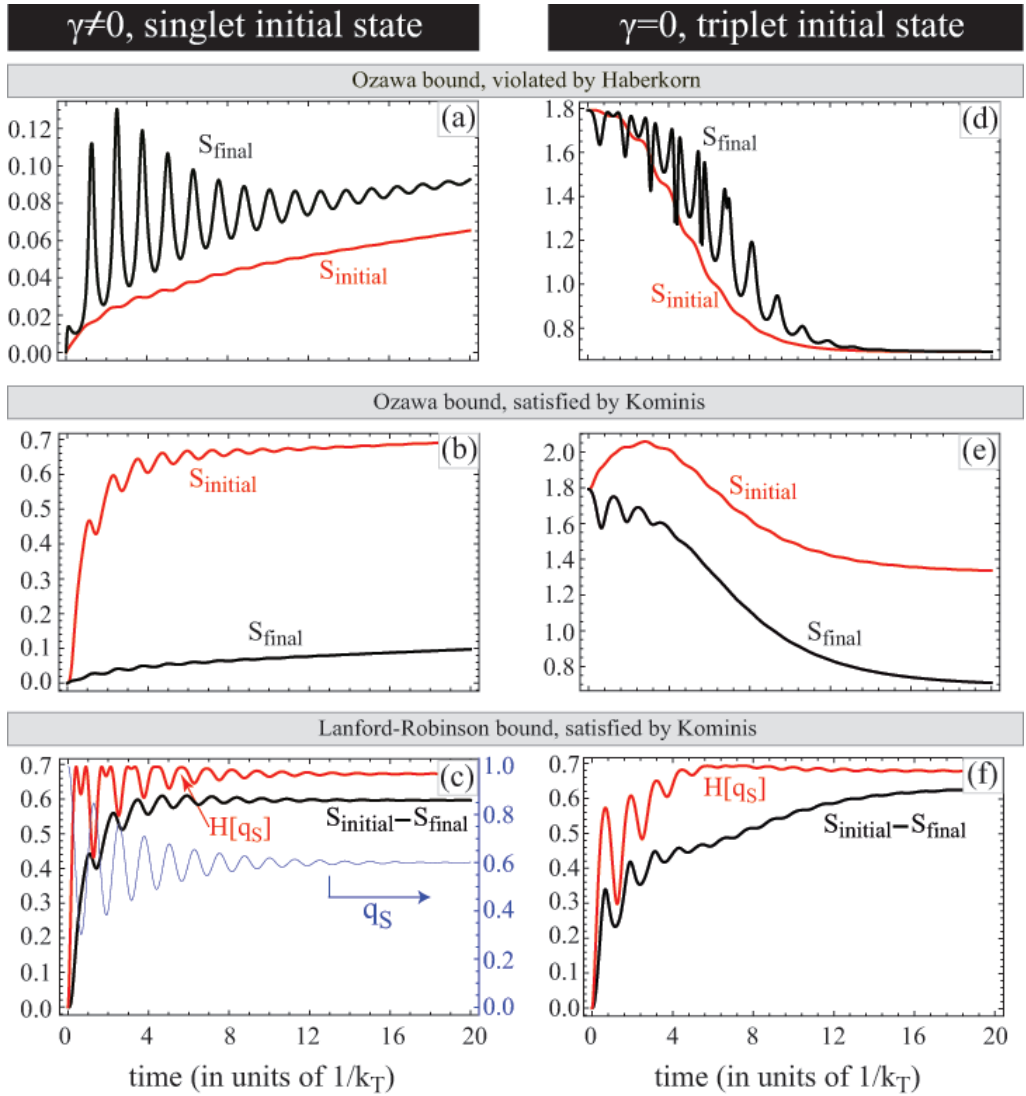}
\caption{The simulations were performed with an isotropic Hamiltonian, ${\cal H}=A\mathbf{I}\cdot\mathbf{s}_{D}$ and asymmetric recombination rates, $\ks=A/100$ and $\kt=A/5$. For (a)-(c) the initial state was $\ket{\rm S}\otimes\ket{\Uparrow}$ and the spin-randomizing term had a rate $\gamma=A/2000$. For (d)-(f) the initial state was the fully mixed triplet, $\qt/{\rm Tr}\{\qt\}$, and there was no spin-randomizing term ($\gamma=0$). The right y-axis in (c) depicts the time-dependence of $q_{\rm S}={\rm Tr}\{\rho\qs\}/{\rm Tr}\{\rho\}$ and demonstrates that the Lanford-Robinson bound is saturated, as expected, at those times when the radical-pair state is close to an eigenstate of the measurement operator $\qs$.}
\label{SDiff}
\end{center}
\end{figure*}

To define the pre- and post-measurement entropies we first note that since the population of the RPs, given by $\tr\{\rho\}$, is time-dependent, with $\tr\{\rho\}=1$ at $t=0$ and $\tr\{\rho\}=0$ at $t\rightarrow\infty$, the single-RP state at time $t$ is $\rho/\tr\{\rho\}$ (see Appendix A for an explanation of the normalization $\rho/\tr\{\rho\}$). Hence, the von-Neumann entropy per RP at any time during the reaction is $S_{\rm initial}={\cal S}\llbracket{\rho/ {\tr\{\rho\}}}\rrbracket$. This form of the initial or pre-measurement entropy is obviously the same for both theories (it is the actual $\rho$ at time $t$ that is different in each theory).
\begin{figure*}[t!]
\begin{center}
\includegraphics[width=11.6 cm]{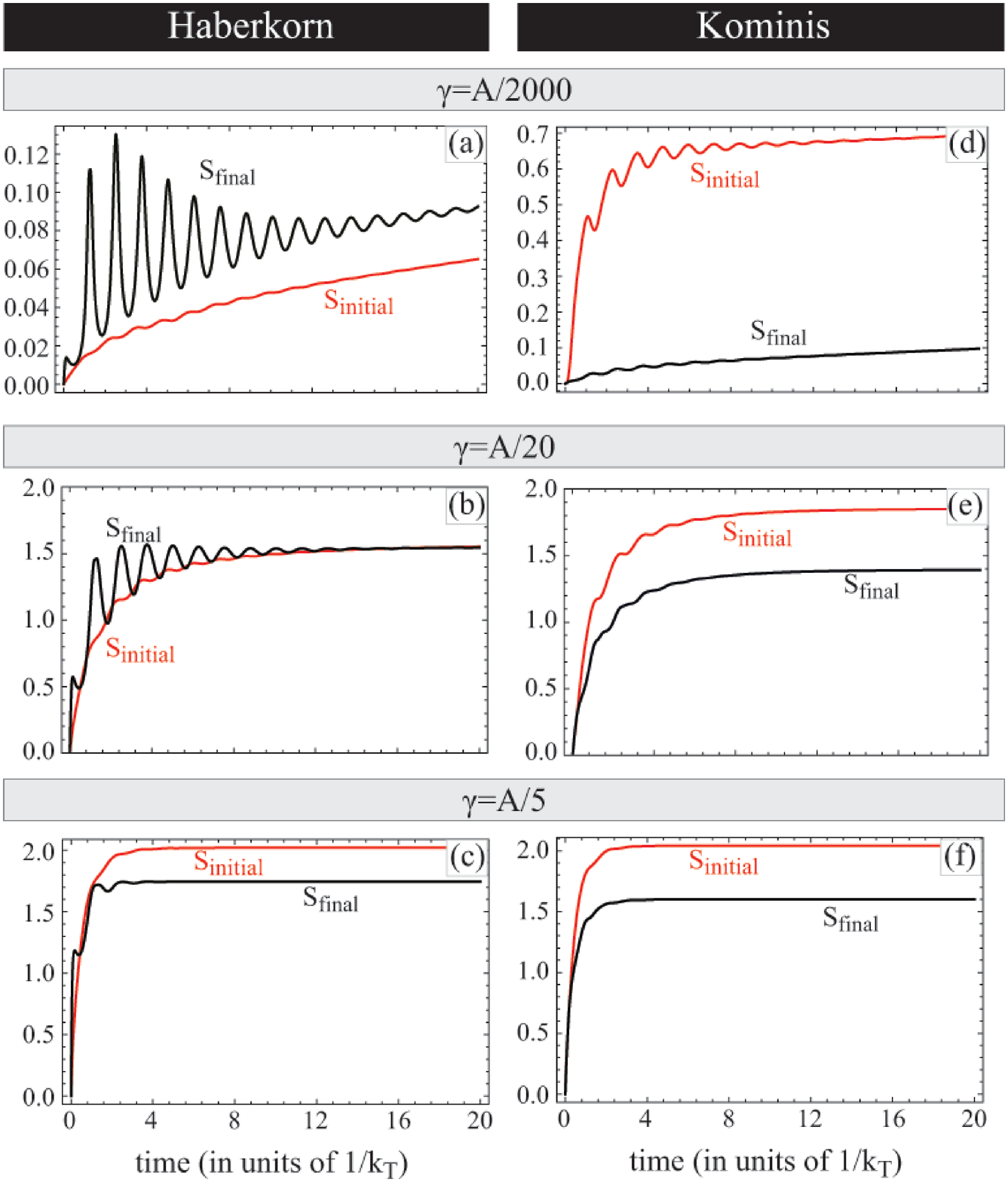}
\caption{For the same Hamiltonian and recombination rates as in Fig. \ref{SDiff}, a pure singlet initial state $\ket{\rm S}\otimes\ket{\Uparrow}$, and for various spin-randomization rates $\gamma$, we plot the initial and final entropies resulting from Haberkorn's (a-c) and Kominis' (d-f) master equation. The first row ($\gamma=A/2000$) reiterates Figs. 1a,b. By increasing $\gamma$ it is seen (b) that Haberkorn's violation first becomes milder, while for still higher $\gamma$ (c) the violation turns into a satisfaction of the bound. This behavior demonstrates that for a large spin relaxation rate $\gamma$, singlet-triplet coherence quantified by $p_{\rm coh}$ swiftly decays to zero, and HME becomes a good approximation describing a mixture of radical-pairs that indeed in incoherent.}
\label{SDiff_gamma}
\end{center}
\end{figure*}

Now, from our theory's perspective, during the time interval $dt$ at time $t$ there will be (i) $dr_{\rm S}$ singlet and $dr_{\rm T}$ triplet neutral products, as well as (ii) $dp_{\rm S}$ and $dp_{\rm T}$ projections (defined in Sec. II) to the singlet and triplet RP states, respectively. The latter also constitute a measurement in the singlet-triplet basis, leading again to the post-measurement states $\rho_{\rm S}=\qs\rho\qs/{\rm Tr}\{\rho\qs\}$ and $\rho_{\rm T}=\qt\rho\qt/{\rm Tr}\{\rho\qt\}$. The only difference is that in case (i) the electron is localized back in the donor, which is irrelevant for the spin state of the molecule described by $\rho_{\rm S}$ and $\rho_{\rm T}$. In other words, during the time interval $dt$ around time $t$, there are $dr_{\rm S}+dp_{\rm S}+dr_{\rm T}+dp_{\rm T}$ radical pairs in the state $\rho/\tr\{\rho\}$, of which $dr_{\rm S}+dp_{\rm S}$ end up in the state $\rho_{\rm S}$, and $dr_{\rm T}+dp_{\rm T}$ end up in the state $\rho_{\rm T}$.
Hence from our theory's perspective, the post-measurement entropy at time $t$ is 
\begin{align}
S_{\rm final}^{\rm K}&={{dr_{\rm S}+dp_{\rm S}}\over {dr_{\rm S}+dr_{\rm T}+dp_{\rm S}+dp_{\rm T}}}{\cal S}\llbracket\rho_{\rm S}\rrbracket\nonumber\\
&+{{dr_{\rm T}+dp_{\rm T}}\over {dr_{\rm S}+dr_{\rm T}+dp_{\rm S}+dp_{\rm T}}}{\cal S}\llbracket\rho_{\rm T}\rrbracket
\end{align}
From Haberkorn's theory perspective, lacking the projections with probability $dp_{\rm S}$ and $dp_{\rm T}$, there are only singlet and triplet neutral products produced during $dt$, hence
\beq
S_{\rm final}^{\rm H}={{dr_{\rm S}}\over {dr_{\rm S}+dr_{\rm T}}}{\cal S}\llbracket\rho_{\rm S}\rrbracket+{{dr_{\rm T}}\over {dr_{\rm S}+dr_{\rm T}}}{\cal S}\llbracket\rho_{\rm T}\rrbracket
\eeq
The reduction of the post-measurement entropy is about the information conveyed by the measurement. According to the Lanford-Robinson bound \cite{Jacobs,Lanford}, this information is bound by the Shannon entropy of the pre-measurement state: $S_{\rm initial}-S_{\rm final}\leq H[q_{\rm S}]$, where $q_{\rm S}=\tr\{\rho\qs\}/\tr\{\rho\}$ and $q_{\rm T}=1-q_{\rm S}$ are the probabilities that the radical-pair is in the singlet or triplet state, respectively, and $H[q_{\rm S}]=-q_{\rm S}{\rm Log}(q_{\rm S})-(1-q_{\rm S}){\rm Log}(1-q_{\rm S})$.
\subsection{Results and Discussion}
In Fig.\ref{SDiff}a and Fig.\ref{SDiff}d we show Haberkorn's prediction for $S_{\rm initial}$ and $S_{\rm final}$, for the case B1 and B2, respectively. Evidently, the Ozawa bound is violated, and hence, the difference $S_{\rm initial}-S_{\rm final}$ being negative, it is meaningless to test the Lanford-Robinson bound. In Figs.\ref{SDiff}b,e we show the result of our master equation, which satisfies the Ozawa bound, and the positive difference $S_{\rm initial}-S_{\rm final}$ satisfies the Lanford-Robinson bound, depicted in  Figs.\ref{SDiff}c,f. 

As a further test of our approach, we note that \cite{Jacobs} the Lanford-Robinson bound is saturated (inequality becomes equality) for semiclassical measurements, i.e. when the measurement operator commutes with the measured system's density matrix. Here the measurement operator is $\qs$, and in the right y-axis of Fig. \ref{SDiff}c we plot the expectation value of $\qs$, $q_{\rm S}={\rm Tr}\{\rho\qs\}/{\rm Tr}\{\rho\}$. It is seen that at the maxima of $q_{\rm S}$, where the radical-pair's spin state is approximately an eigenstate of $\qs$, the entropy difference approaches Shannon's information $H[q_{\rm S}]$.

To understand the root of Haberkorn's violation, we omit the rate $\ks$, since anyhow we consider the case $\kt\gg\ks$. It follows from HME that the coherence $\qs\rho\qt+\qt\rho\qs$ decays at the rate $\kt/2$, whereas the population $\qt\rho\qt$ decays at the rate $\kt$. Thus the coherence decays just due to the population loss, i.e. there is no intrinsic dissipation of coherence, and therefore an initially pure state remains pure (apart from the spin-randomizing term). In contrast, from our theory it follows that the coherence is dissipated at the rate $\kt[1/2+ q_{\rm T}]\geq \kt/2$, whereas population decays at the rate $\kt[1+p_{\rm coh}(q_{\rm T}-1)]\leq\kt$ (obviously  $0\leq q_{\rm T}\leq 1$). Thus the coherence decay is faster than what would result just due to population loss, which makes for entropy production.

On a more abstract level, HME fails to account for the fact that RP recombination essentially is a rate process {\it conditioned} on the quantum state of the molecule. To understand this subtle point consider two-level atoms (with two long-lived states $\ket{g}$ and $\ket{e}$) escaping a box with a hole, {\it if they are in the $\ket{g}$ state}. The hole's diameter determines the escape rate (the equivalent of the recombination rates). However, every time the atom approaches the hole, some physical process must measure the atom's state, which in general could be in any coherent superposition of $\ket{g}$ and $\ket{e}$. If the result of this measurement is positive, the atom will escape with a given probability. It is this measurement (the equivalent of the state projections leading to S-T dephasing in our theory) that is an entropy source for the atoms remaining in the box. Measurement and escape are two independent processes, which HME attempts to treat as one. 

To understand how a small randomization rate can produce the observed effect on the HME entropy (Fig. \ref{SDiff}a) that would otherwise be zero, we note that according to the scenario of Fig. \ref{SDiff}a, we start initially from a pure state, having zero entropy, as correctly depicted in Fig. \ref{SDiff}a. At time $t=t_{\rm max}=20/\kt=100/A$ it is $\gamma t_{\rm max}=0.05$, hence the state at that time is "almost" pure, with an admixture of a fully mixed state having a 5\% weight. The entropy of the fully mixed state is Log[8] (since we are considering an 8-dimensional density matrix), and 0.05Log[8]=0.10, which is roughly equal to the y-axis values of Fig. \ref{SDiff}a at $t=t_{\rm max}$.

Finally, in Fig. \ref{SDiff_gamma} we demonstrate the transition from a violation to a satisfaction of the Ozawa bound by Haberkorn's theory, a transition that takes place by increasing the spin-randomization rate $\gamma$. In Figs. \ref{SDiff_gamma}a,d we reiterate Figs. \ref{SDiff}a,b, i.e. the case $\gamma=A/2000$. We then increase $\gamma$. In Fig. \ref{SDiff_gamma}b it is seen that Haberkorn's violation becomes less pronounced, i.e. $S_{\rm final}$ approaches $S_{\rm initial}$ from above, while for an even larger $\gamma$ (Fig. \ref{SDiff_gamma}c), Haberkorn's theory satisfies the Ozawa inequality. As mentioned earlier, Haberkorn's theory cannot properly capture singlet-triplet coherence, because (at low or zero $\gamma$) it attempts to describe the dynamics of S-T coherent RPs treating them as incoherent. However, in the presence of high enough spin relaxation (large $\gamma$), S-T coherence is anyhow damped much faster than its intrinsic damping rate $(\ks+\kt)/2$, hence the radical-pair spin state approaches an S-T incoherent mixture, i.e. it is described by $p_{\rm coh}=0$. In this regime HME offers a satisfactory approximation of the underlying dynamics. Returning to the biological context of the radical-pair mechanism, it is a matter of what kind and how strong are relaxation mechanisms at play in the chemical system under consideration that will disqualify the use of HME and require another theory, like \eqref{t1}-\eqref{t4}. 
\section{New magnetic-field-effect predicted from Groenewold information}
Besides serving as a test of the master equation, the quantum information approach to RP reactions is useful in its own right. This is because, by definition, it carries in an abstract way the full information that can be extracted from the reaction by {\it any kind} of measurement. Based on the form of $S_{\rm final}^{\rm K}$, we define 
\begin{align}
{\cal I}_{\rm G}=\int&\Big[(dr_{\rm S}+dr_{\rm T}+dp_{\rm S}+dp_{\rm T})S_{\rm initial}\nonumber\\
&-(dr_{\rm S}+dp_{\rm S}){\cal S}\llbracket\rho_{\rm S}\rrbracket-(dr_{\rm T}+dp_{\rm T}){\cal S}\llbracket\rho_{\rm T}\rrbracket\Big]\label{IG}
\end{align}
as the integrated extracted information. This is known as the Groenewold information, the utility of which we demonstrate next. 

\begin{figure}[t!]
\begin{center}
\includegraphics[width=8.5 cm]{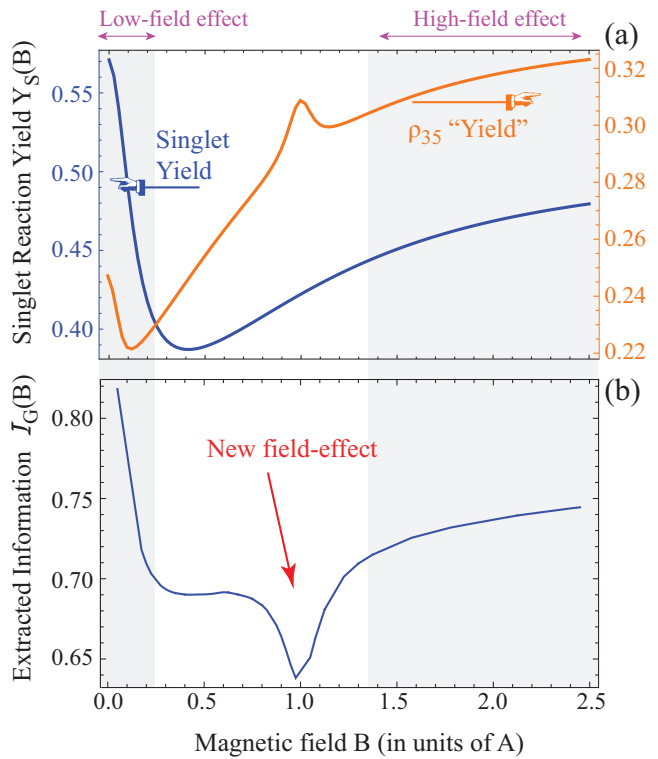}
\caption{(a) Singlet reaction yield dependence (left y-axis) on the magnetic field using \eqref{t1}-\eqref{t4}, depicting low- and high-field effects. (b) Extracted (Groenewold) information ${\cal I}_{\rm G}(B)$, calculated from \eqref{IG}, exhibits both low- and high-field effect, but also reveals a new field-effect at $B=A$. This is evidenced by "measuring" the coherence $\rho_{35}$, the relevant "yield" shown in the right y-axis of (a). The calculation was performed with ${\cal H}=A\mathbf{I}\cdot\mathbf{s}_{D}+B(s_{Dz}+s_{Az})$, $\ks=\kt=A/20$,  initial state $\ket{\rm S}\otimes\ket{\Uparrow}$, and no spin-randomizing term ($\gamma=0$).}
\label{info}
\end{center}
\end{figure}

A central observable in RP reactions is the magnetic field effect \cite{Rodgers}, i.e. the reaction yield dependence on the applied magnetic field $B$. For example, considering the singlet reaction yield, $Y_{\rm S}=\int dr_{\rm S}$, the dependence $Y_{\rm S}(B)$ stems from the modulation of the S-T mixing by the Zeeman terms $B(s_{Dz}+s_{Az})$ of the donor and acceptor electron entering the Hamiltonian. In Fig.\ref{info}a we plot (left y-axis) the ordinary field effect $Y_{\rm S}(B)$, depicting the low-field effect due to zero-field level crossings, and the high-field effect due to the triplet states shifting out of resonance with the singlet \cite{Timmel}. In Fig.\ref{info}b we plot ${\cal I}_{\rm G}(B)$. Apparently, this quantity also conveys the low-field and the high-field effect. That is, ${\cal I}_{\rm G}(B)$ has a steep $B$-dependence at low $B$ and a milder $B$-dependence at high $B$, just like $Y_{\rm S}(B)$. Interestingly though, it carries {\it additional} information, as evidenced by the dip at $B=A$. 

The reason that this new field-effect is not observed in the singlet reaction yield of Fig.\ref{info}a  is that measuring the singlet character of the RP state is not optimal for extracting this particular field-effect. By changing the measurement, and instead of the projector $\qs$, "measuring" the absolute value of the element $\rho_{35}$ of the density matrix we can observe such a field effect, as seen in Fig.\ref{info}a (right y-axis). This matrix element corresponds to the coherence $\ket{\uparrow\downarrow}\bra{\downarrow\uparrow}\otimes\ket{\Uparrow}\bra{\Uparrow}$. Thus, the utility of ${\cal I}_{\rm G}(B)$ is to reveal the magnetic sensitivity of the reaction imprinted in {\it any kind of measurement}. 

How to perform realistic generalized measurements \cite{Caves} optimally extracting the information that can in principle be extracted will be explored elsewhere, along with the origin of the new field effect. In particular, understanding the latter, i.e. the field effect apparent in the $\rho_{35}$-"yield", requires the involved exercise of expressing the non-reacting evolution law $d\rho/dt=-i[{\cal H},\rho]-(\ks+\kt)(\rho\qs+\qs\rho-2\qs\rho\qs)/2$ in Liouville space, $d\tilde{\rho}/dt={\cal K}\tilde{\rho}$, where $\tilde{\rho}$ is a column vector formed by joining all columns of $\rho$.  The eigenvalues of ${\cal K}$ are $-\lambda_m+i\Omega_m$, with $\lambda_m\geq 0$, and hence the matrix elements of $\rho$ are given by $\rho_{ij}=\sum_{m}c_{ij}^{(m)}e^{-\lambda_{m}t+i\Omega_{m}t}$, where $c_{ij}^{(m)}$ depend on the initial state. The decay rate $\lambda_m$ is a function of $B$, therefrom stems
a $B$-dependence of the lifetime and hence the contribution of particular combinations of $\rho_{ij}$ in the measured observable. 
\section{Conclusions}
In summary, we introduced the quantum information and entropy perspective in radical-pair reactions, which are a central paradigm in the new field of quantum biology. This approach serves as a sharp test for our understanding of the fundamental quantum dynamics of the radical-pair mechanism, since it is not plagued by debatable comparisons of absolute theoretical predictions with each other or with experimental data. Instead, our approach is based on fundamental entropy inequalities, the violation of which 
can unambiguously rule out the master equation traditionally used until now in treating radical-pair reactions. 

Furthermore, entropy considerations further support our previous quantum trajectory analysis \cite{cpl2015}. For example, considering the entropy of RPs along the lines of \cite{Comment_Jeschke} is not even possible in the first place, since the density matrices introduced in \cite{Comment_Jeschke} in an attempt to consistently describe RP quantum trajectories from Haberkorn's perspective have negative eigenvalues \cite{Reply_Kominis}. In other words, both this work and the quantum trajectory analysis of \cite{cpl2015} pose to the contenders of Haberkorn's approach the challenge to provide a picture of the radical-pair dynamics at the single-molecule level that together with the ensemble description of the density matrix via the master equation lead to a consistent description satisfying the entropy bounds presented here.

Introducing the concept of Groenewold information extracted from the reaction leads to a deeper insight of the metrological aspect of RP reactions, revealing new magnetic-field effects conveyed in a general and abstract way, i.e. without reference to any particular measurement scheme.

Finally, we provide further evidence that the master equation we have developed captures the quantum dynamics of RP reactions, since it is shown to satisfy the Ozawa bound as well as the Lanford-Robinson bound, saturating the latter at times when 
it is expected to be saturated on physical grounds.
\appendix
\section{}
To explain why we define the single-RP entropy using $\rho/\tr\{\rho\}$ as the single-RP state, consider for example a fully mixed radical-pair spin state with population $p<1$. It is described by a diagonal density matrix $\rho=(p/8)\mathbbmtt{1}$, where $\mathbbmtt{1}$ is the 8-dimensional unit matrix, so that indeed $\tr\{\rho\}=p$. Since the state is fully mixed, we expect its entropy to be Log[8]. Indeed, normalizing $\rho$ by $\tr\{\rho\}$, we find ${\cal S}\llbracket\rho/\tr\{\rho\}\rrbracket=Log[8]$. If instead we calculate ${\cal S}\llbracket\rho\rrbracket$, we get a nonlinear and non-sensical function of $p$, ${\cal S}\llbracket\rho\rrbracket=-p{\rm Log}[p/8]$. If we were to normalize this expression by $p$,  the result would still not be a sensible entropy. Hence the former is the correct way to normalize the density matrix. This same normalization, $\rho/\tr\{\rho\}$, has also been used elsewhere, see e.g. \cite{Jones}.


\begin{thebibliography}{10}
\bibitem{basis}
In this work we will consider an RP having just one nuclear spin, hence the density matrix is 8-dimensional. The 8 basis kets are $\ket{\rm S}\otimes\ket{\Uparrow}$, $\ket{\rm S}\otimes\ket{\Downarrow}$, $\ket{\rm T_+}\otimes\ket{\Uparrow}$, $\ket{\rm T_+}\otimes\ket{\Downarrow}$, $\ket{\rm T_0}\otimes\ket{\Uparrow}$, $\ket{\rm T_0}\otimes\ket{\Downarrow}$, $\ket{\rm T_-}\otimes\ket{\Uparrow}$, $\ket{\rm T_-}\otimes\ket{\Downarrow}$. The two-electron state is on the left of the tensor product and the nuclear spin state on the right. The former is spanned by the singlet $\ket{\rm S}=(\ket{\uparrow\downarrow}-\ket{\downarrow\uparrow})/\sqrt{2}$ and the triplets $\ket{\rm T_+}=\ket{\uparrow\uparrow}$, $\ket{\rm T_0}=(\ket{\uparrow\downarrow}+\ket{\downarrow\uparrow})/\sqrt{2}$, $\ket{\rm T_-}=\ket{\downarrow\downarrow}$. Although obvious, we mention that the RP model we choose is of no concern, since we here demonstrate a violation of a fundamental law of entropy by the foundational theory of spin chemistry. As well known, it takes infinite examples to establish a theory, but just one counter-example to invalidate it.
\end{thebibliography}
\end{document}